\documentclass[prb,twocolumn,showpacs,superscriptaddress]{revtex4}
\usepackage[latin1]{inputenc}
\usepackage{graphicx}
\usepackage{amssymb}
\usepackage{amsmath}
\usepackage{xspace}
\usepackage{dcolumn}
\usepackage{bm}
\usepackage{color}
\usepackage[extra]{tipa}

\newfont{\tensy}{cmsy10}

\newcommand{\ie}[0]{i.e.\@\xspace}
\newcommand{\eg}[0]{e.g.\@\xspace}
\newcommand{\etal}[0]{{\em et al.}\@\xspace}

\newcommand{\UP}[0]{\uparrow}
\newcommand{\DO}[0]{\downarrow}

\newcommand{\oh}{\mbox{$\frac{1}{2}$}}

\newcommand{\si}[0]{\sigma}

\newcommand{\nag}{{\phantom{\dag}}}

\newcommand{\Dsp}{\Delta_\text{sp}}
\newcommand{\Ds}{\Delta_\text{s}}
\newcommand{\Sxy}{S_\text{AF}^{xy}}
\newcommand{\Szz}{S_\text{AF}^{zz}}

\newcommand{\Uc}{U_\text{c}}
\newcommand{\lc}{\lambda_\text{c}}

\newcommand{\las}[0]{\langle}
\newcommand{\ras}[0]{\rangle}
\newcommand{\llas}[0]{\langle\langle}
\newcommand{\rras}[0]{\rangle\rangle}
\newcommand{\la}[0]{\left\las}
\newcommand{\ra}[0]{\right\ras}
\newcommand{\ket}[1]{\left|#1\ra}
\newcommand{\bra}[1]{\la#1\right|}

\newcommand{\Tr}[0]{\mbox{Tr}}

\begin{document}


\title{Quantum phase transitions in the Kane-Mele-Hubbard model}

\author{M. Hohenadler}
\affiliation{\mbox{Institut f\"ur Theoretische Physik und Astrophysik,
    Universit\"at W\"urzburg, 97074 W\"urzburg, Germany}}

\author{Z. Y. Meng}
\affiliation{Center for Computation and Technology, Louisiana State
  University, Baton Rouge, Louisiana 70803, USA} 
\affiliation{Institut f\"ur Theoretische Physik III, Universit\"at Stuttgart,
  70550 Stuttgart, Germany}

\author{T. C. Lang}
\affiliation{\mbox{Institute for Theoretical Solid State Physics, JARA-FIT,
 and JARA-HPC, RWTH Aachen University, 52056 Aachen, Germany}}

\author{S. Wessel}
\affiliation{\mbox{Institute for Theoretical Solid State Physics, JARA-FIT,
 and JARA-HPC, RWTH Aachen University, 52056 Aachen, Germany}}

\author{A. Muramatsu}
\affiliation{Institut f\"ur Theoretische Physik III, Universit\"at Stuttgart,
  70550 Stuttgart, Germany}

\author{F. F. Assaad}
\affiliation{\mbox{Institut f\"ur Theoretische Physik und Astrophysik,
    Universit\"at W\"urzburg, 97074 W\"urzburg, Germany}}

\begin{abstract}
  We study the two-dimensional Kane-Mele-Hubbard model at half filling by means of quantum
  Monte Carlo simulations. We present a refined phase boundary for the
  quantum spin liquid. The topological insulator at finite Hubbard
  interaction strength is adiabatically connected to the groundstate
  of the Kane-Mele model. In the presence of spin-orbit coupling, magnetic
  order at large Hubbard $U$ is restricted to the transverse direction.
  The transition from the topological band insulator to the antiferromagnetic
  Mott insulator is in the universality class of the three-dimensional XY
  model. The numerical data suggest that the spin liquid to topological
  insulator and spin liquid to Mott insulator transitions are both continuous.
\end{abstract}

\date{\today}


\maketitle

\section{Introduction}\label{sec:intro}

Topological insulators have attracted significant attention in recent
years,\cite{HaKa10} especially since their experimental
realization.\cite{Koenig07} Whereas the existence of the topological 
state and many of its consequences can be understood in terms of exactly
solvable, noninteracting models, the interplay of a topological band
structure and electronic correlations has become a very active field
of research. The corresponding interacting models do not have general 
exact solutions, which has made computational methods one of the most
important tools. A possible experimental route to the strongly correlated
regime is based on optical lattices.\cite{PhysRevLett.107.145301}

The $Z_2$ topological band insulator (TBI), or quantum spin-Hall insulator,
closely related to the integer quantum Hall effect,\cite{HaKa10} can be realized in the
Kane-Mele (KM) model.\cite{KaMe05a,KaMe05b} The latter describes
electrons (or Dirac fermions) on the two-dimensional (2D) honeycomb lattice, with
nearest-neighbor hopping and spin-orbit coupling. Originally motivated
by graphene,\cite{KaMe05b} the spin-orbit coupling turned out to be much too
small in this material for topological effects to be observable. However, the KM model and its extension,
the Kane-Mele-Hubbard (KMH) model turn out to be a very useful theoretical
framework. In particular, the honeycomb lattice geometry provides a direct
connection to the recently discovered quantum spin liquid (QSL) phase of the
Hubbard model on the same lattice.\cite{Meng10} For the latter, the Dirac
spectrum with vanishing density of states at the Fermi level leads to a Mott
transition at a finite critical Hubbard $U$, and the QSL phase lies between a
semimetal and a magnetic insulator.\cite{Meng10} Finally, the 
symmetries of the KMH model permit the application of powerful quantum Monte
Carlo (QMC) methods without a sign problem,\cite{Hohenadler10,Zh.Wu.Zh.11} so
that exact results can be obtained.

The phase diagram of the KMH model has been derived from QMC simulations,
\cite{Hohenadler10,Zh.Wu.Zh.11} and numerical results for
the extent of the QSL were presented in
Ref.~\onlinecite{Hohenadler10}. At any nonzero spin-orbit coupling, the
semimetal is replaced by the $Z_2$ TBI. In contrast, the gapped
QSL is found to be stable up to a finite critical value of the spin-orbit
interaction. Finally, the magnetic transition of the Hubbard model, between
the QSL and an antiferromagnetic Mott insulator (AFMI), is supplemented with a similar
transition between the TBI and the AFMI in a potentially different
universality class. On a qualitative level, certain aspects of the phase diagram
were obtained for example in mean-field theory,\cite{RaHu10} as well as with
cluster methods\cite{Yu.Xie.Li.11,Wu.Ra.Li.LH.11} and variational
QMC.\cite{PhysRevB.83.205122}

The understanding of the KMH model is not complete. Many of the open
questions are related to the perhaps most intriguing aspect of the model,
namely the QSL phase. The recent results from approximate cluster methods for
parameters  in the QSL region of the exact phase diagram
inaccurately suggest a rather complete understanding of this exotic
phase. However, strictly speaking, any cluster method breaks translational
symmetry, so that a true QSL phase is excluded from the outset. In this
light, conclusions such as the absence of edge states, or the closing of the
single-particle gap across the transition to the TBI are not surprising,
as the QSL phase is replaced in these studies by a simple band insulator (a valence bond
crystal). The large correlation lengths (small gaps) observed
in the QSL phase in the Hubbard model\cite{Meng10}
highlight the necessity of careful interpretation of the results obtained by
cluster approximations in the context of the QSL. Interesting connections between TBIs
and QSLs are discussed in Ref.~\onlinecite{Fi.Ch.Hu.Ka.Lu.Ru.Zy.11}. 

The purpose of this paper is threefold. First, we present a much more
detailed account of the QMC calculations underlying the phase diagram shown
in Ref.~\onlinecite{Hohenadler10}. Second, we extend the number of points in
parameter space and the observables calculated, in order to provide
additional insight. We also present a refined phase boundary
for the QSL phase. Third, we use the QMC method to investigate the 
quantum phase transitions, especially in the light of recent theoretical
predictions.\cite{PhysRevLett.107.166806,Gr.Xu.11} We show that the TBI--AFMI
transition is in the expected 3D XY universality class, and provide evidence
for the continuous nature of the QSL--TBI and the QSL--AFMI quantum phase transitions. In
contrast to earlier work,\cite{Hohenadler10} we only consider bulk
properties. We also provide an overview of recent
work on correlation effects in topological insulators with a focus on the KMH model.

The paper is organized as follows. In Sec.~\ref{sec:model} we briefly review
the model. Details about the QMC method are presented in
Sec.~\ref{sec:method}. Section~\ref{sec:results} contains our numerical
results, beginning with the refined phase diagram, and followed by a detailed
account of the various quantum phase transitions. We end with conclusions and
an overview of open questions in Sec.~\ref{sec:conclusions}.

\section{Model}\label{sec:model}

The Hamiltonian of the KMH model can be written in the form ${H=H_\text{KM}+H_U}$, where
\begin{align}\label{eq:H}
  H_\text{KM} &= -t \sum_{\las \bm{i},\bm{j} \ras}
  c^{\dagger}_{\bm{i}} c^\nag_{\bm{j}} + i\,\lambda \sum_{\llas\bm{i},\bm{j}\rras} 
  \nu^\nag_{\bm{i}\bm{j}}  c^{\dagger}_{\bm{i}} \sigma^z  c^\nag_{\bm{j}} \,,\nonumber \\
  H_U &= \frac{U}{2} \sum_{\bm{i}} (c^{\dagger}_{\bm{i}} c^\nag_{\bm{i}} -
  1 )^2\,.
\end{align}
Here ${c^{\dagger}_{\bm{i}} = \big(c^{\dagger}_{\bm{i},\uparrow},
  c^{\dagger}_{\bm{i},\downarrow}\big)}$ is a spinor of electron creation
operators, $\bm{i}$ is the position of a lattice site on the honeycomb
lattice, $\las\bm{i},\bm{j}\ras$ denotes a pair of nearest
neighbors, and $\llas \bm{i},\bm{j} \rras$ is a
pair of next-nearest-neighbor lattice sites; $\sigma^z$ is a Pauli matrix,
and $\nu_{\bm{ij}}=\pm 1$ depending on whether the hopping path defined by
the nearest-neighbor bonds connecting sites $\bm{i}$ and $\bm{j}$ bends
to the right or to the left. The complex
next-nearest-neighbor hopping term in Eq.~(\ref{eq:H}) can be related to the
spin-orbit interaction in graphene and accounts for a spin-dependent staggered
magnetic field.\cite{KaMe05b} The choice of writing the
interaction term $H_U$ in an $SU(2)$ invariant form is related to previous work on
the Hubbard model on the honeycomb lattice,\cite{Meng10} in which it was
essential to build this symmetry into the QMC method, see also
Sec.~\ref{sec:method}.  In the presence of the spin-orbit term,
the $SU(2)$ spin rotation symmetry is reduced to a $U(1)$ symmetry.
Throughout this work, we use periodic boundary
conditions so that there are no edges, and take $t$ as the unit of
energy. The number of unit cells  in each direction is denoted by $L$, the
total number of unit cells is $L^2$, and the total number of lattice sites is $N=2L^2$. The lattice
sizes used satisfy $L=3l$ with $l$ integer, and range from $L=3$ to $L=18$.
We exclusively consider the case of a half-filled band.

A possible Rashba term is neglected from the outset, because it would cause a
sign problem in the QMC simulations. However, a small but finite Rashba coupling
does not destroy the TBI state of the KM model.\cite{KaMe05a} The
noninteracting case $U=0$ has been solved in the original paper by Kane and
Mele.\cite{KaMe05a} Most importantly, the groundstate is a $Z_2$ TBI for any
finite spin-orbit coupling $\lambda$. The KM model is closely related to a
spinless model proposed by Haldane which shows a quantum Hall effect and
breaks time reversal invariance (TRI).\cite{Ha88} Combining two copies of the
Haldane model gives the KM model exhibiting the quantum spin-Hall effect and preserving TRI.\cite{KaMe05a,Fi.Ch.Hu.Ka.Lu.Ru.Zy.11}

Hamiltonian~(\ref{eq:H}) has been studied by means of mean-field and analytical
approaches,\cite{RaHu10,PhysRevB.82.161302,JPSJ.80.044707,We.Ka.Va.Fi.11,Va.Ma.Ho.11}
QMC simulations,\cite{Hohenadler10,Zh.Wu.Zh.11,PhysRevB.83.205122} the
variational cluster approach,\cite{Yu.Xie.Li.11} cluster dynamical mean-field
theory,\cite{Wu.Ra.Li.LH.11} and field theory.\cite{PhysRevLett.107.166806,Gr.Xu.11} A more
detailed discussion of previous results will be given in
Sec.~\ref{sec:results}.

\section{Method}\label{sec:method}

We employ a projective auxiliary-field determinant QMC algorithm similar to
Ref.~\onlinecite{Meng10}, which has previously been applied to the KMH
model.\cite{Hohenadler10,Zh.Wu.Zh.11} The method is based on the relation
\begin{equation}
  \label{qmc.eq}
  \langle
  \Psi_0 | O |\Psi_0 \rangle = \lim_{\theta \rightarrow \infty} 
  \frac{
    \langle
    \Psi_\text{T} | e^{-\theta H/2} O e^{-\theta H/2} |
    \Psi_\text{T} \rangle} {\langle \Psi_\text{T} | e^{-\theta H} |
    \Psi_\text{T} 
    \rangle}
\end{equation}
for the expectation value of an operator $O$, with a trial wave function
$|\Psi_\text{T} \rangle $ that is required to be
nonorthogonal to the groundstate $\ket{\Psi_0}$.  It is beyond the scope of this article
to describe the details of the algorithm, and the interested reader is
referred to Ref.~\onlinecite{AsEv08}. Instead, we  concentrate on aspects specific
to the calculations presented here, namely the choice of the trial
wave function, the Hubbard-Stratonovich (HS) transformation,
and the absence of the minus-sign problem for a half-filled
band.

\subsection{Trial wave function}

In order to simplify the implementation, the trial wave
function is taken to be a single Slater determinant, and can hence always
be written in terms of the groundstate of a single-particle Hamiltonian
$H_\text{T}$. There are many possible choices for $\ket{\Psi_\text{T}}$. One
can for example decide to optimize the overlap with the groundstate at the
expense of symmetries.\cite{Furukawa91} Here we have preserved symmetries, and
have chosen $|\Psi_\text{T} \rangle $ to be the groundstate of the KM model, 
which is defined by the first line of Eq.~(\ref{eq:H}).  For $\lambda\neq 0$, the
groundstate of the noninteracting problem at half filling is
insulating. Hence, the trial wave function is nondegenerate and has all the
symmetries of the Hamiltonian.  At $\lambda = 0$, the situation is more
delicate. For the considered lattice sizes, $ L = 3l$, the two nonequivalent
Dirac points are located at the Fermi surface, and the groundstate of the
noninteracting model at half filling is four-fold degenerate in each spin
sector.  We lift this degeneracy by means of a twist in the boundary
condition in $H_\text{T}$ in the direction $\bm{a}_1=(1,0)$,
\begin{equation}
  H_\text{T} = -t\sum_{\langle \bm{i},\bm{j} \rangle } c^{\dagger}_{\bm{i}}c^{\nag}_{\bm{j}} 
  \,\exp \left[\frac{ 2 \pi i }{ \Phi_0}  \int_{\bm{i}}^{\bm{j}}  \bm{A} \cdot d \bm{l} \right]\,,
\end{equation}
with $\bm{A} = \Phi \bm{a}_1/L $. The twist preserves translation
symmetry, so that the total momentum remains a good quantum number.  In
particular, for an infinitesimal twist, the groundstate has vanishing total
momentum. Because finite values of $\Phi$ lead to a breaking of the $C_3$
lattice symmetry, the trial wave function cannot be classified according to the
irreducible representation of this group at $\lambda=0$.
After lifting possible degeneracies, $ | \Psi_\text{T} \rangle $
corresponds to the nondegenerate groundstate of $H_\text{T}$. This implies
the relation
\begin{equation}\label{eq:proj}
  \lim_{\Theta \rightarrow \infty }    e^{-\Theta (H_\text{T}  - E_\text{T})}    
  =  |  \Psi_\text{T} \rangle \langle \Psi_\text{T} |\,,
\end{equation}
where $E_\text{T}$ is the corresponding groundstate energy.

\subsection{Hubbard-Stratonovich transformation}

We choose a HS transformation of the Hubbard term $H_U$  that couples
to the total density $n_i=n_{i\UP}+n_{i\DO}$, thereby conserving the $SU(2)$ spin symmetry for every
field configuration.  In principle, HS transformations that couple to the
$z$-component of spin are also possible. However, at low temperatures, it is
often difficult to restore the $\lambda=0$ $SU(2)$ spin symmetry of the total Hamiltonian
by stochastic sampling. An $SU(2)$ spin symmetric
transformation was previously used in Refs.~\onlinecite{Assaad99,Capponi00}.  

After a Trotter decomposition with imaginary time step $\Delta \tau$ to isolate
the interaction term, our HS transformation for a general operator $O$ reads
\begin{equation}\label{eq:HS}
  \ensuremath e^{-\Delta\tau O^2} =
  \sum_{l=\pm 1,\pm 2} \gamma(l)
  \ensuremath {e}^{i\sqrt{\Delta\tau }\, \eta(l)  O } + \mathcal{O}(\Delta\tau^{4}) \,,
\end{equation} 
with the two functions $\gamma(l)$ and $\eta(l)$ of the auxiliary field
$l$ (with $l = \pm 1,\pm 2$) taking on the values
\begin{align}
  \gamma(\pm 1) = (1+\sqrt{6}/3)/4 \,,&\quad& \eta(\pm 1) = \pm\sqrt{2\,(3-\sqrt{6})} \,,\nonumber\\
  \gamma(\pm 2) = (1-\sqrt{6}/3)/4 \,,&\quad& \eta(\pm 2) = \pm\sqrt{2\,(3+\sqrt{6})} \,.
\end{align}
Equation~(\ref{eq:HS}) is an approximation to the Gaussian integral and
introduces an overall systematic error of the order
$\Delta\tau^{3}$, which is negligible in comparison to the
Trotter error of order $\Delta\tau^{2}$.  The major advantage of
this approximation is that we can avoid using continuous auxiliary fields
while retaining spin rotation symmetry.
For the Hubbard interaction, we have $O = \sqrt{U/2}(n_{\uparrow}+ n_{\downarrow} -
1)$ and it is understood that the HS fields acquire space and time indices,
$l\mapsto l_{\bm{i},\tau}$.
%

\subsection{Absence of a sign problem}

We prove the absence of the minus-sign problem for the {\it projective} QMC
method at half filling. With the Trotter decomposition, choice of trial wave function and
HS transformation, the denominator of Eq.~(\ref{qmc.eq}) factors into spin-up
and spin-down determinants, 
\begin{widetext}
\begin{equation}
  \langle \Psi_\text{T}  | \prod_{\tau=1}^{L_\tau} e^{-\Delta \tau
    H_\text{KM} } 
  e^{-\Delta \tau H_U} | \Psi_\text{T} \rangle  
  =
  \Tr \left[
    \lim_{\Theta\rightarrow \infty} e^{-\Theta (H_\text{T}  - E_\text{T})}    
     \prod_{\tau=1}^{L_\tau} e^{-\Delta \tau  H_\text{KM} } e^{-\Delta \tau H_U}
  \right]
  =  
  \lim_{\Theta\rightarrow \infty} 
  \sum_{ \{ l_{i,\tau} \}}\prod_{\sigma}
  \prod_{\tau=1}^{L_\tau} \prod_{\bm{i}} \gamma(l_{\bm{i},\tau})  W_{\sigma}\,,
\end{equation}
where we have used Eq.~(\ref{eq:proj}) to introduce a trace,
$\{l_{\bm{i},\tau}\}$ denotes an auxiliary-field configuration, and with the
weights
\begin{equation}\label{eq:Wsigma}
  \begin{split}
    W_{\sigma}  =  
    \Tr \bigg[  
      e^{\Theta E_\text{T}}
      \exp\bigg\{
      -\Theta 
      \sum_{\bm{ij}} c^{\dagger}_{\bm{i}\si}
      [h^\nag_\text{T}(\Phi)]^\nag_{\bm{ij}}  c^\nag_{\bm{j}\si}  
      \bigg\}
      \prod_{\tau=1}^{L_{\tau}}  
      &\exp\bigg\{
      -\Delta \tau \sum_{\bm{ij}} c^{\dagger}_{\bm{i}\si}  
      \left( A_t + A_{\lambda,\sigma} \right)_{\bm{ij}} c^\nag_{\bm{j}\si}
      \bigg\}
      \\ 
      &\times\exp\bigg\{i \sqrt{\Delta \tau U/2} \sum_{\bm{i}} \eta(l_{\bm{i},\tau})(n_{\bm{i}\si} - 1/2) \bigg\}
      \bigg]\,. 
  \end{split}
\end{equation}
Here we introduced the notation $ H_\text{T} = \sum_{\bm{ij}\sigma}c^{\dagger}_{\bm{i}\sigma}
[h^\nag_\text{T}(\Phi)]^\nag_{\bm{ij}} c^\nag_{\bm{j}\sigma} $ and $H_\text{KM} =\sum_{\bm{ij}\sigma}c^{\dagger}_{\bm{i}\sigma}
( A_t + A_{\lambda,\sigma} )^\nag_{\bm{ij}} c^\nag_{\bm{j}\sigma} $.
Proving the absence of a negative sign problem at half
filling amounts to showing that $W_{\uparrow}^*=W_{\downarrow} $. Since the trace in
Eq.~(\ref{eq:Wsigma}) is over one spin sector, we drop the spin index
on the fermion operators to lighten the notation and obtain
\begin{align}  
  W^*_{\uparrow}  
  &= 
  \Tr \bigg[ 
   e^{\Theta E_\text{T}}
  \exp\bigg\{
  -\Theta 
  \sum_{\bm{ij}} c^{\dagger}_{\bm{i}} [h^*_\text{T}(\Phi)]^\nag_{\bm{ij}}  c^\nag_{\bm{j}}  \bigg\}
  \prod_{\tau=1}^{L_{\tau}} 
  \exp\bigg\{
  -\Delta \tau  \sum_{ij} c^{\dagger}_{\bm{i}} 
  ( A^*_t + A^*_{\lambda,\uparrow} )^\nag_{\bm{ij}} c^\nag_{\bm{j}}
  \bigg\}
  \\\nonumber
  & \qquad\qquad \qquad\qquad \qquad\qquad \qquad\qquad  
  \qquad\qquad \nonumber
  {} \times\exp\bigg\{
  -i \sqrt{\Delta \tau U/2} \sum_{\bm{i}} \eta(l_{\bm{i},\tau}) (n_{\bm{i}} -1/2) 
  \bigg\} 
  \bigg]  
  \\ 
  &=\nonumber
  \Tr \bigg[   
  e^{\Theta E_\text{T}}
  \exp\bigg\{
  -\Theta 
  \sum_{\bm{ij}} c^\nag_{\bm{i}} [h_\text{T}^*(\Phi)]^\nag_{\bm{ij}}(-1)^{\bm{i}+\bm{j}}  c^{\dagger}_{\bm{j}} 
  \bigg\}
  \prod_{\tau=1}^{L_{\tau}} 
  \exp\bigg\{-\Delta \tau  \sum_{\bm{ij}} c^{\nag}_{\bm{i}} 
  ( A^*_t + A^*_{\lambda,\uparrow} )^\nag_{\bm{ij}} (-1)^{\bm{i}+\bm{j}}
  c^{\dagger}_{\bm{j}}  \bigg\}  
  \\\nonumber
  & \qquad\qquad \qquad\qquad \qquad\qquad \qquad\qquad  \qquad\qquad \qquad\qquad
  {} \times
  \exp\bigg\{ i \sqrt{\Delta \tau U/2} \sum_{\bm{i}}
    \eta(l_{\bm{i},\tau})( 1 - n_{\bm{i}} - 1/2) \bigg\}     
  \bigg]
  \,.
\end{align}
%
%
\end{widetext}
The second line follows from the {\it canonical} transformation $c^\nag_{\bm{i}}
\rightarrow (-1)^{\bm{i}} c^{\dagger}_{\bm{i}} $, where the phase factor $(-1)^{\bm{i}}$
takes the value $1$ ($-1$) on sublattice A (B).  The Hamiltonian $H_\text{T}$
which generates the trial wave function has nonvanishing matrix elements only
between sites on opposite sublattices. Hence $ (-1)^{\bm{i}+\bm{j}}= -1$ and
\begin{equation}\label{eq:sign1}
  c^\nag_{\bm{i}} [h^*_\text{T}(\Phi)]^\nag_{\bm{ij}}(-1)^{\bm{i}+\bm{j}} c^{\dagger}_{\bm{j}} 
  =
  c^{\dagger}_{\bm{j}}  [h^*_\text{T}(\Phi)]^\nag_{\bm{ij}}  c^\nag_{\bm{i}} 
  = 
  c^{\dagger}_{\bm{j}}  [h_\text{T}^\nag(\Phi)]^\nag_{\bm{ji}} c^\nag_{\bm{i}}.
\end{equation}
Similarly, for the hopping term,
\begin{equation}\label{eq:sign2}
  c^\nag_{\bm{i}}  ( A^*_t )^\nag_{\bm{ij}}
  (-1)^{\bm{i}+\bm{j}} c^{\dagger}_{\bm{j}}  
  = 
  c^{\dagger}_{\bm{j}}    \left( A_t\right)^\nag_{\bm{ji}}  c^\nag_{\bm{i}} .
\end{equation}
Since the spin-orbit term involves hopping between sites on the same
sublattice, we have
\begin{eqnarray}\label{eq:sign3}\nonumber
  c^\nag_{\bm{i}} 
  ( A^*_{\lambda,\uparrow} )^\nag_{\bm{ij}} (-1)^{\bm{i}+\bm{j}}
  c^{\dagger}_{\bm{j}}  
  &= 
  -c^{\dagger}_{\bm{j}}  \left( A_{\lambda,\uparrow} \right)^\nag_{\bm{ji}} c^\nag_{\bm{i}}   \\
  &= 
  c^{\dagger}_{\bm{j}}  \left( A_{\lambda,\downarrow} \right)^\nag_{\bm{ji}} c^\nag_{\bm{i}}\,.
\end{eqnarray}
Using Eqs.~(\ref{eq:sign1})--(\ref{eq:sign3}), one sees that indeed
\begin{equation}
  W^*_{\uparrow} = W_{\downarrow}\,, 
\end{equation}
so that no sign problem exists at the particle-hole symmetric point of
the KMH model in the present formulation of the QMC algorithm.\cite{Hohenadler10}
The underlying reason is time reversal symmetry, which implies
$A_{\lambda,\downarrow} = - A_{\lambda,\uparrow} $.

\subsection{Measurements}

For a given auxiliary-field configuration, we have to solve a free-electron Hamiltonian
with external fields that vary in time and space. Consequently, Wick's
theorem holds, and it is sufficient to compute the single-particle Green functions
\begin{equation}
  G_{\sigma}(\bm{i},\bm{j},\tau,\tau') =  
  - \bra{\Psi_0} T c^\nag_{\bm{i},\sigma}(\tau) c^\dag _{\bm{j},\sigma}(\tau')  \ket{\Psi_0}
\end{equation}
to calculate arbitrary correlation functions. For the calculation of $G_\sigma$ we have followed
Ref.~\onlinecite{Feldbach00}. The single-particle gap $\Dsp$ at the Dirac
point and the (staggered) spin gap $\Ds$ at $\bm{q}=0$ are extracted from
fits to the corresponding Green functions.\cite{Meng10}

\subsection{Projection parameter and Trotter discretization}

The projective algorithm involves two numerical parameters, namely the
projection parameter $\theta$ and the Trotter time step $\Delta\tau$.
Both parameters were chosen such that their influence on the results is
smaller than the statistical errors. Explicitly, we used $\Delta\tau
  t=0.05$ or 0.1, and $\theta t=40$--$60$.

\section{Results}\label{sec:results}

In order to better orient the discussion, we first present the phase diagram
of the KMH model in Sec.~\ref{sec:pd} together with a review of recent work,
before elaborating on the various quantum phase transitions.

\begin{figure}[t]
  \includegraphics[width=0.45\textwidth]{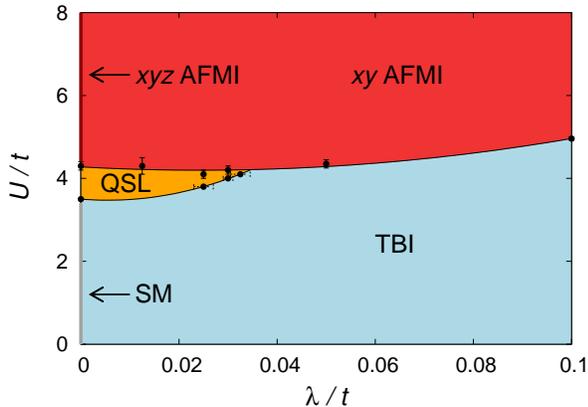}
  \caption{\label{fig:pd} (Color online) Groundstate phase diagram of the
    Kane-Mele-Hubbard model as obtained from QMC simulations. The
    four phases are a $Z_2$ topological band insulator (TBI) with nonzero
    single-particle (spin) gap $\Dsp>0$ ($\Ds>0$), a
    semimetal (SM, $\Dsp=\Ds=0$) existing at $\lambda=0$, a quantum spin liquid
    (QSL, $\Dsp>0$, $\Ds>0$), and an antiferromagnetic Mott insulator (AFMI, $\Dsp>0$,
    $\Ds=0$). Magnetic order in the $z$ direction exists in the AFMI at $\lambda=0$,
    but can be excluded for $\lambda/t\geq0.002$ and all values of $U/t$ shown.
    Lines are quadratic fits to the QMC data points.
    }
\end{figure}

\subsection{Phase diagram}\label{sec:pd}

Figure~\ref{fig:pd} shows the groundstate phase diagram of the KMH model, as
obtained from QMC simulations. In addition to the three phases of the Hubbard
model on the honeycomb lattice, the spin-orbit coupling introduces a $Z_2$ TBI. The
gapless SM phase exists only at $\lambda=0$.  Whereas the TBI and the QSL
phase are fully gapped (finite single-particle gap $\Dsp$ and spin
gap $\Ds$), the magnetic phase has $\Dsp>0$ but
$\Ds=0$.  Here all gaps refer to the bulk, and are not to be
confused with the metallic, gapless edge states of the TBI phase of the KMH model.

To the best of our knowledge, the QSL phase is characterized by the absence
of any local order parameter which would reflect a broken-symmetry state. It can hence be regarded as a
genuine Mott insulating state, which should be stable with respect to
small perturbations such as spin-orbit coupling. In the case of an odd number of electrons per
 unit cell, the generalization of the Lieb-Schultz-Mattis theorem to two
dimensions\cite{Hasting04} suggests the presence of topological order in the
most general sense.  Since
the half-filled honeycomb lattice has two electrons per unit cell, this topological
ordering still has to be numerically demonstrated or refuted.  The
underlying $SU(2)\times SU(2)/Z_2$ symmetry of the Hubbard model on the
honeycomb lattice has led to the prediction of a $Z_2\times Z_2$ QSL
with mutual spin-charge statistics.\cite{PhysRevB.83.024408} Sublattice pairing states have been put
forward by various authors to account for the QSL
phase.\cite{PhysRevLett.107.087204,PhysRevB.84.024420} A canonical
consequence of the above topologically ordered phases is that, assuming
a continuous phase transition, the magnetically ordered phase would
not be a simple N\'{e}el state, thus leading to conjectures that can be tested
numerically.\cite{PhysRevLett.107.087204,PhysRevB.84.024420} Finally, in the presence of
spin-orbit coupling, the possibility of the emergence of a topological
Mott insulating phase, in which the spinons carry the topological character of
the phase, remains.\cite{PeBa10,Ru.Fi.11} For the KMH model, recent theoretical
suggestions include a $Z_2$ QSL\cite{Gr.Xu.11} and a chiral QSL.\cite{Va.Ma.Ho.11}

The boundary of the magnetic phase is obtained from the onset of long-range
antiferromagnetic order in the $xy$ plane. Longitudinal order, present at
$\lambda=0$, can be excluded in Fig.~\ref{fig:pd} for all $U/t$ and for
$\lambda/t\geq0.002$, so that the $xyz$ AFMI phase is confined to a very 
small (possibly infinitesimal) interval starting at $\lambda=0$. The SM--TBI transition is evinced by the
simultaneous opening of a single-particle and a spin gap, which
as a function of $\lambda$ closely follow the $U=0$ results. The QSL--TBI transition for
intermediate Hubbard $U$ and small $\lambda$ turns out to be the most
difficult and perhaps most interesting case, with the critical values
extracted from a cusp (consistent with a closing) of the single-particle gap
$\Dsp$ and the spin gap $\Ds$. A more detailed discussion is given below.

Our numerical results suggest that the TBI phase at finite $U$ is
adiabatically connected to the TBI state of the KM model ($U=0$).  Similarly, the
QSL phase is stable over a finite range of $\lambda$, in
accordance with theoretical predictions.\cite{Gr.Xu.11} Except for the
smaller range of spin-orbit couplings compared to Ref.~\onlinecite{Hohenadler10}, which is chosen here to
highlight the  structure of the phase diagram around the QSL, we have obtained a number of additional points for the
phase boundary of the QSL. The refined QSL phase boundary reveals a
direct magnetic transition between the QSL and the AFMI phase at finite
$\lambda$. Our numerical data suggest the existence of a
multicritical point where the QSL, TBI and AFMI phases meet. The estimated
location of this point is  $(\lc,\Uc)\approx(0.035t,4.2t)$.

Let us compare the phase diagram in Fig.~\ref{fig:pd} to other work. The
magnetic phase boundary was calculated using mean-field
theory.\cite{RaHu10} In that work, a transition from the TBI to an AFMI
phase is observed, with the critical $U$ increasing with increasing
$\lambda$ and comparable to the band width. However, the numerical values differ by up to a
factor of two.  The phase diagram from unbiased QMC simulations was
presented by three of us.\cite{Hohenadler10} At that time, only one
point on the QSL--TBI phase boundary was available, and the suggested
dome-like structure of the QSL phase was based on the fact that the spin gap
takes on its maximum around $U/t=4$, in the middle of the $\lambda=0$ QSL
phase.   Soon after this work, QMC results for the phase diagram
were published by Zheng \etal\cite{Zh.Wu.Zh.11} Except for the absence of
the QSL--TBI phase boundary, their phase diagram is compatible with
previous\cite{Hohenadler10} and current results
(Fig.~\ref{fig:pd}).  The line $\lambda/t=0.1$ was studied by Yamaji and
Imada using variational QMC simulations,\cite{PhysRevB.83.205122} although
with rather large quantitative differences concerning the location of the
TBI--AFMI transition. The phase diagram has also been
calculated using the variational cluster approach,\cite{Yu.Xie.Li.11} and
cluster dynamical mean-field theory.\cite{Wu.Ra.Li.LH.11} Apart from the fact
that a true QSL phase is not accessible in any cluster calculation, the overall
structure of the phase diagram in these works is consistent with Fig.~\ref{fig:pd}.
The quantitative phase boundaries seem to be slightly more accurate
in the cluster dynamical mean-field case.\cite{Wu.Ra.Li.LH.11} Both papers
show a ``QSL''--TBI phase boundary whose shape is in accordance with our refined
phase diagram in Fig.~\ref{fig:pd}. 

Lee\cite{PhysRevLett.107.166806} and Griset and Xu\cite{Gr.Xu.11} have
recently made predictions about the nature of some of the phase
transitions. In both works, the TBI--AFMI transition is
argued to be in the 3D XY universality class, as already hinted at in
Ref.~\onlinecite{Hohenadler10}.  Griset and Xu
further suggest that both the QSL--AFMI and the QSL--TBI transitions could be
first order quantum phase transitions.\cite{Gr.Xu.11} They also highlight
the possibility of an additional, nematic order-disorder transition inside
the AFMI at $\lambda=0$, instead of a proposed chiral AF order-disorder
transition in the Hubbard model\cite{PhysRevB.84.024420} that should persist
also at $\lambda>0$.\cite{Gr.Xu.11} The phase diagram of the KMH model has
also been calculated using analytical methods.\cite{We.Ka.Va.Fi.11,Va.Ma.Ho.11}

With the number of phases and their boundaries being rather well established,
the important open questions about the phase diagram concern the nature of
the QSL and AFMI phases, and of the various phase transitions. The
structure of the phase diagram implies the existence of several distinct
quantum phase transitions: SM--TBI, TBI--AFMI, QSL--AFMI, and QSL--TBI. The
remaining SM--QSL transition only occurs at $\lambda=0$, and has been studied in
detail before.\cite{Meng10} We discuss each of these transitions below.

The remainder of this section is organized as follows. We first consider
the SM--TBI transition (at fixed $U/t=2$, see Fig.~\ref{fig:pd}) and the
TBI--AFMI transition (at fixed $\lambda/t=0.1$), for which we can
provide a fairly complete picture. From this we move on to the QSL--AFMI
transition (considering $\lambda/t=0.0125$), and finally the QSL--TBI
transition (at $U/t=4$).

\subsection{Semimetal to topological insulator transition}\label{sec:sm-tbi}

We begin with the SM--TBI transition. To this end, we keep $U/t=2$ fixed.  In
the absence of interactions, the  spin-orbit term breaks the sublattice
symmetry and generates a mass gap as well as a topological band structure.  Due to the
underlying $U(1)$ spin symmetry, the band structure corresponds to two
Haldane models with Chern numbers of opposite sign in the two spin
sectors.\cite{KaMe05a} 

\begin{figure}[t]
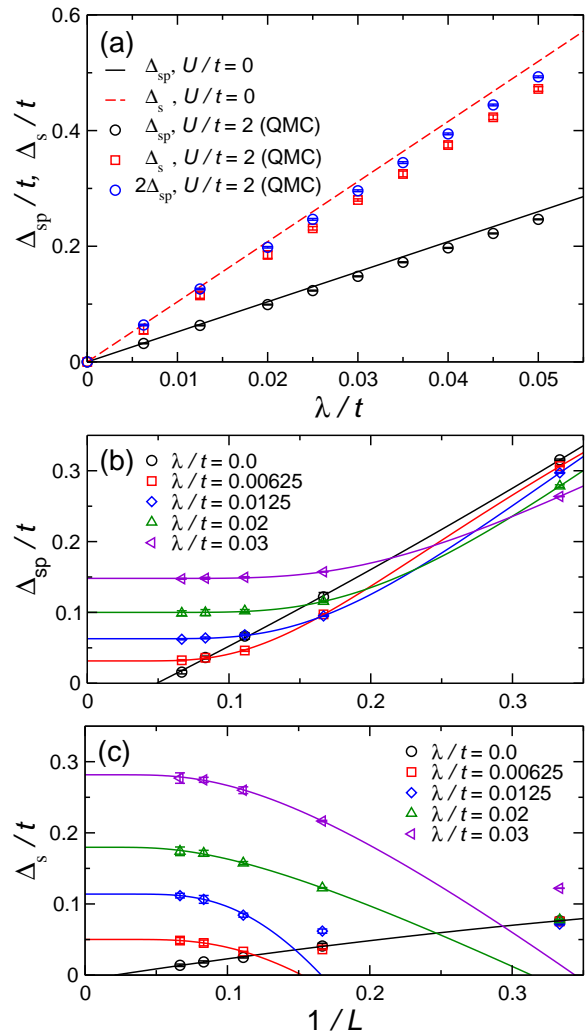

  \includegraphics[width=0.425\textwidth]{fig_sm-tbi}\vspace*{0.5em}  
  \includegraphics[width=0.425\textwidth]{fig_sm-tbi_sp}\vspace*{0.5em}
  \includegraphics[width=0.425\textwidth]{fig_sm-tbi_spin}
  \caption{\label{fig:sm-tbi} (Color online) (a) Single-particle gap $\Dsp$,
    $2\Dsp$ and spin gap $\Ds$ as a function of $\lambda$ at $U/t=2$, across
    the SM--TBI transition. The QMC results are obtained from finite-size
    extrapolation using the fitting function Eq.~(\ref{eq:fit2}) at $\lambda=0$
    and Eq.~(\ref{eq:fit1}) for $\lambda>0$. The numerical results shown as symbols agree
    well with the corresponding gaps of the KM model ($U=0$),\cite{KaMe05b}
    $\Dsp=3\sqrt{3}\lambda$ and $\Ds=2\Dsp$, revealing the adiabatic
    connection between the TBI at $U=0$ and at $U>0$.  (b) Finite-size
    scaling of the single-particle gap at selected  values of $\lambda/t$. 
    (c) Finite-size scaling of the spin gap; for $\lambda>0$, we neglect
    the $L=3$ results in the extrapolation.}
\end{figure}

Figure~\ref{fig:sm-tbi}(a) shows QMC results for the
single-particle gap $\Dsp$ and the spin gap $\Ds$ as a function of
$\lambda$ at $U/t=2$.  Starting in the SM
phase at $\lambda=0$, where both gaps are zero, $\Dsp$ and $\Ds$ become nonzero for any finite
$\lambda$, and increase with increasing spin-orbit coupling. The QMC results
at $U/t=2$ closely follow the corresponding gaps
for the noninteracting case $U=0$,\cite{KaMe05b} $\Dsp=3\sqrt{3}\lambda$ and
$\Ds=2\Dsp$. Interaction effects manifest themselves as a minor suppression
of both gaps compared to their noninteracting values, especially at
larger $\lambda/t$, and by the spin gap falling below $2\Dsp$.

From these results, we draw the following conclusions. First, the SM phase of
the Hubbard model is unstable at finite $\lambda$, and hence only exists for
$\lambda=0$, as indicated in Fig.~\ref{fig:pd}. Second, the very small
deviations in the dependence of the gaps on $\lambda$ compared to the
noninteracting case suggest that the TBI phase at $U>0$ is essentially the
same as at $U=0$, provided $U$ remains small enough to avoid the magnetic
transition. This finding suggests that the two states are adiabatically
connected. The minor role of bulk interactions inside the TBI phase may
be regarded as a consequence of the single-particle energy gap,\cite{HaKa10}
and has been exploited to develop an effective model of the helical edges
with a Hubbard $U$ only at the edge sites of a ribbon.\cite{Hohenadler10,Ho.As.11}

The results for $\Dsp$ and $\Ds$ in Fig.~\ref{fig:sm-tbi}(a) are obtained
from finite-size scaling, as shown for selected values of $\lambda/t$ in
Figs.~\ref{fig:sm-tbi}(b) and (c). Whereas $\Dsp$ shows the familiar
monotonic decrease with increasing system size, the spin gap reveals an
unusual finite-size scaling behavior. Deep in the TBI phase [\eg, the top
curve in Fig.~\ref{fig:sm-tbi}(c) corresponding to $\lambda/t=0.03$],
$\Ds$ systematically increases with increasing system size $L$. For small
$\lambda$ (for example, $\lambda/t=0.0125$) and small $L$, the scaling
behavior is more complex, and only the system sizes beyond the crossover have
been used in the extrapolation. The increase of $\Ds$ with increasing $L$
is a correlation effect; the spin gap is independent of $L$ for $U=0$.
The observed increase of $\Ds$ with system size can be reproduced using first-order
perturbation theory in $U$. 

A possible physical explanation is inspired by the observation that for small $L$,
$\Ds(L)< 2\Dsp(L)$, see Figs.~\ref{fig:sm-tbi}(b) and (c). In contrast, the
extrapolated values almost match the relation for the noninteracting case,
$\Ds\approx 2\Dsp$, as shown in Fig.~\ref{fig:sm-tbi}(a). The strongly
suppressed spin gap for small system sizes indicates pronounced particle-hole
binding, driven by correlation-induced magnetic fluctuations. If the length
scale of these fluctuations, which are a precursor of the magnetic transition
at $\Uc$, exceeds the system size, the spin gap is expected to be suppressed
similar to the magnetic phase where $\Ds\to0$ as $L\to\infty$. Increasing $L$
beyond the correlation length will restore the behavior expected for the
weakly or noninteracting TBI phase. We will see below
(Fig.~\ref{fig:qsl-tbi-sp}) that for larger $U/t=4$, the spin gap is
suppressed to values much below $2\Dsp$ even in the thermodynamic limit.
Although a complete understanding of this effect is currently missing,
we regard the unusual spin gap scaling as a signature of a correlated TBI.

\subsection{Topological insulator to antiferromagnet transition}\label{sec:tbi-afmi}

At large $U/t$, the TBI phase of the half-filled KMH model undergoes a
transition to an AFMI, and TRI is spontaneously broken.\cite{RaHu10}
In the strong-coupling limit, $U / t \gg 1$, the charge degrees of freedom are frozen
and one can derive an effective spin model with antiferromagnetic nearest-neighbor Heisenberg
exchange $J=4t^2/U$ that promotes isotropic magnetic order in the $xy$ and
$z$ directions. The spin-orbit term of the KMH model reduces the $SU(2)$ spin symmetry of the
Hubbard model to a $U(1)$ symmetry corresponding to conservation of the total
$z$-component of spin. Second-order perturbation theory gives an exchange
interaction $J'=4\lambda^2/U$ between next-nearest
neighbors. Importantly, the exchange is antiferromagnetic in the longitudinal
direction, $J' S^z_{\bm{i}} S^z_{\bm{j}}$, but ferromagnetic in the
transverse direction, $-J'(S^x_{\bm{i}} S^x_{\bm{j}}+S^y_{\bm{i}} S^y_{\bm{j}})$.\cite{RaHu10} 
Combining all the exchange terms, magnetic order in the $z$ direction becomes
frustrated, and the system favors an easy-plane N\'eel state. The so-called
KM-Heisenberg model was recently studied analytically.\cite{Va.Ma.Ho.11}

\begin{figure}[t]
  \includegraphics[width=0.425\textwidth]{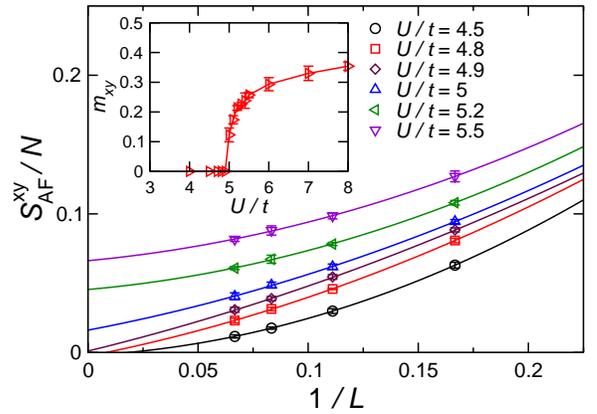}
  \caption{ \label{fig:tbi-afmi} (Color online) 
    Finite-size scaling of the rescaled magnetic structure factor
    $\Sxy/N$ defined in Eq.~(\ref{eq:SAF}) at
    $\lambda/t=0.1$ for different values of $U/t$, across the TBI--AFMI
    transition.  The curves (representing polynomial fits) extrapolate to zero for
    $U/t<4.9$, and to a finite value for $U/t\geq5.0$, giving the critical value $\Uc/t=4.95(5)$.
    A more accurate estimate $\Uc/t=4.96(4)$ is obtained from
    Fig.~\ref{fig:3DXY_SAFxy}(a). The inset shows the order parameter $m^{xy}$ as obtained from
    extrapolation to the thermodynamic limit.}
\end{figure}

From the above considerations, $xy$ order is expected both for $\lambda=0$
and $\lambda\neq0$. Hence, the phase boundary of the AFMI phase can be
determined from the onset of transverse long-range magnetic order by
monitoring the transverse structure factor
\begin{align}\label{eq:SAF}
 \Sxy
 &\equiv \sum_{\alpha} [\Sxy]^{\alpha\alpha}\,,
 \\\nonumber
 [\Sxy]^{\alpha\beta}
 &= \frac{1}{L^2}\sum_{\bm{r}\bm{r}'} (-1)^{\alpha} (-1)^{\beta} 
 \langle \Psi_0
 | S^{+}_{\bm{r}\alpha} S^{-}_{\bm{r}'\beta} + S^{-}_{\bm{r}\alpha} S^{+}_{\bm{r'}\beta} | \Psi_0 \rangle\,.
\end{align}
Here $\bm{r},\bm{r}'$ denote unit cells, $\alpha,\beta\in\{A,B\}$ are sublattice indices,
 $(-1)^\alpha=1$ ($-1$) for $\alpha=A$ ($B$), and we have taken the trace of
  the corresponding $2\times2$ matrix of the structure factor.
The quantity $\Sxy /N$ (with $N=2L^2$) extrapolates to zero below
$\Uc(\lambda)$, but takes on a finite value in the thermodynamic
limit for $U\geq\Uc(\lambda)$, which is inside the AFMI phase of
Fig.~\ref{fig:pd}.\cite{Hohenadler10,Zh.Wu.Zh.11} It is also related
to the transverse magnetization via $m_{xy}^2=\Sxy /N$.

\begin{figure}[t]
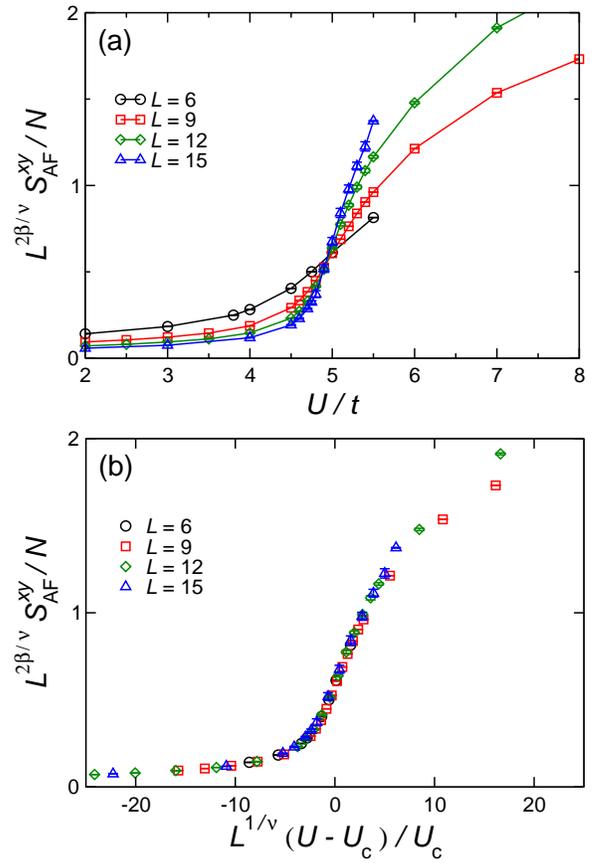

  \includegraphics[width=0.425\textwidth]{fig_tbi-afmi_safxy_intersect}\vspace*{0.5em}
  \includegraphics[width=0.425\textwidth]{fig_tbi-afmi_safxy_collapse}
  \caption{\label{fig:3DXY_SAFxy} (Color online) 
    Rescaled transverse magnetic structure factor $\Sxy/N$
    defined in Eq.~(\ref{eq:SAF}) as a function of $U$ at $\lambda/t=0.1$,
    for different lattice sizes $L$. Assuming the scaling
    form~(\ref{eq:scaling:S}), (a) shows $L^{2\beta/\nu}\Sxy/N$. The intersection of
    curves for different system sizes yields $\Uc/t=4.96(4)$ for the critical point.
    (b) The scaling collapse obtained by plotting  $L^{2\beta/\nu}
    \Sxy/N$ as a function of $L^{1/\nu}
    (U-\Uc)/\Uc$. The QMC data are fully consistent with the critical
    exponents $z=1$, $\nu=0.6717(1)$ and $\beta=0.3486(1)$ of the 3D XY
    model.\cite{PhysRevB.74.144506} }
\end{figure}

Numerical results for $\lambda/t=0.1$ are shown in Fig.~\ref{fig:tbi-afmi};
the transition is most obvious from the extrapolated order
parameter shown in the inset. The extrapolation of
$\Sxy /N$ in system size gives a critical value $\Uc/t=4.95(5)$. This value
agrees with the slightly more accurate estimate $\Uc/t=4.96(4)$ which follows
from the intersect of curves for different system sizes in Fig.~\ref{fig:3DXY_SAFxy}(a).
However, this scaling analysis (see below for more details) relies on the knowledge
of the universality class of the transition. By performing calculations at
different $\lambda/t$, we can determine the magnetic phase boundary, and we
find good agreement with previous exact
simulations at $\lambda=0$\cite{Meng10} and
$\lambda>0$.\cite{Hohenadler10,Zh.Wu.Zh.11} Our QMC results (not
shown) further exclude the presence of longitudinal magnetic order along the
entire TBI--AFMI phase boundary in Fig.~\ref{fig:pd} and up to $U/t=8$.

\begin{figure}[t]
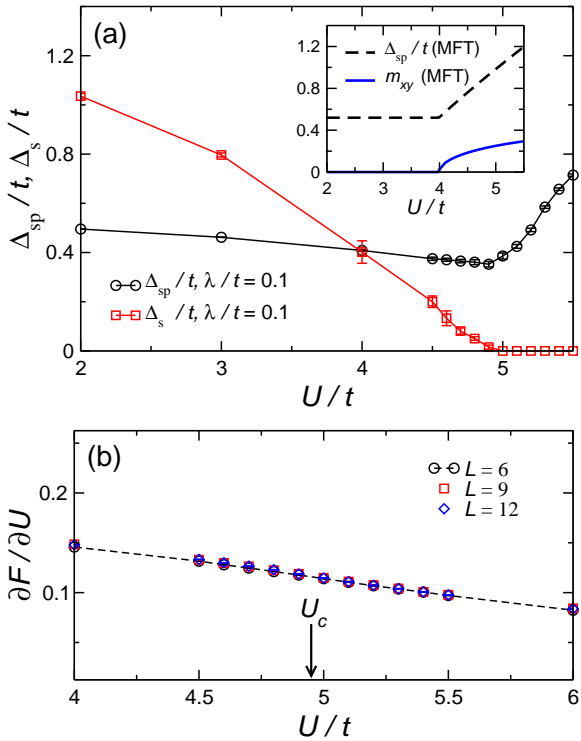

  \includegraphics[width=0.425\textwidth]{fig_tbi-afmi_gaps}\vspace*{0.75em}
  \includegraphics[width=0.425\textwidth]{fig_dfdu_lambda0.1.eps}
  \caption{\label{fig:tbi-afm-gaps} (Color online) (a) Single-particle gap
    $\Dsp$ and spin gap $\Ds$ as a function of $U$ at
    $\lambda/t=0.1$. The values shown were obtained from an extrapolation to
    the thermodynamic limit. The dip in $\Dsp$ and the closing of
    $\Ds$ are consistent with $\Uc/t=4.95(5)$. The inset shows  
    the mean-field results for the single-particle gap and the magnetic
    order parameter. (b) Energy derivative with respect to $U$
    [Eq.~(\ref{eq:dfdU})] across the TBI--AFMI transition at
    $\lambda/t=0.1$ [$\Uc/t=4.95(5)$]. }
\end{figure}

\begin{figure}[t]
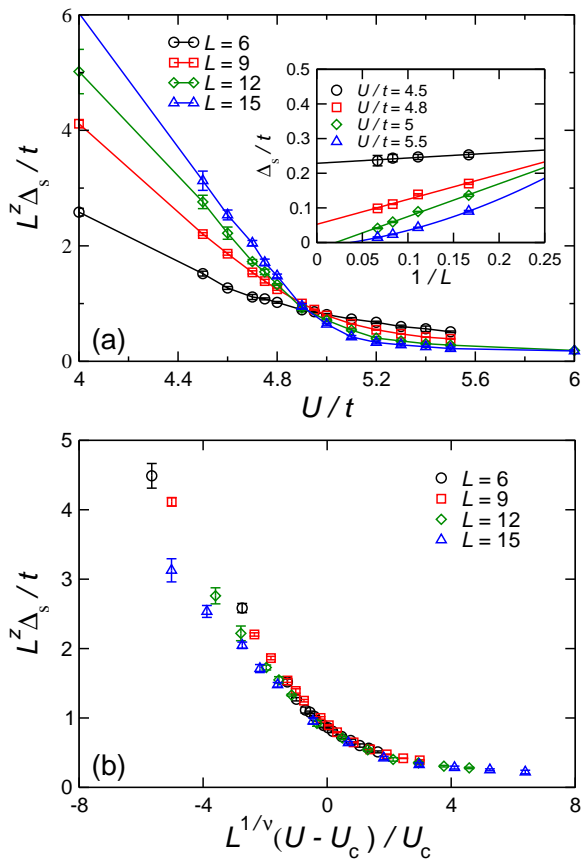

  \includegraphics[width=0.425\textwidth]{fig_tbi-afmi_ds_intersect}\vspace*{0.45em}
  \includegraphics[width=0.425\textwidth]{fig_tbi-afmi_ds_collapse}
  \caption{\label{fig:3DXY_gap} (Color online) 
    Spin gap $\Ds$ as a function of $U$ at
    $\lambda/t=0.1$, for different lattice sizes $L$. Given the scaling
    form Eq.~(\ref{eq:scaling:D}), (a) shows $L^{z} \Ds$. The intersection
    of curves for different $L$ gives $\Uc/t=4.96(4)$, consistent
    with Fig.~\ref{fig:3DXY_SAFxy}(a). The inset shows the finite-size
    scaling of $\Ds$. (b) Scaling collapse obtained by plotting  $L^{z}\Ds$ as a
    function of $L^{1/\nu}(U-\Uc)/\Uc$. The QMC data are consistent with the 3D
    XY exponents $z=1$, $\nu=0.6717(1)$ and
    $\beta=0.3486(1)$.\cite{PhysRevB.74.144506}    
  }
\end{figure}

The TBI--AFMI transition is also reflected in the single-particle gap.
Because the onset of long-range magnetic order at the TBI--AFMI transition
spontaneously breaks TRI, the transition from the TBI to the
nonadiabatically connected AFMI can in principle occur without
closing any excitation gaps. Instead, the transition manifests itself in
$\Dsp$ as a cusp at $\Uc/t=4.95(5)$, visible in Fig.~\ref{fig:tbi-afm-gaps}(a). The
results are for the same value of $\lambda/t=0.1$ considered in Fig.~\ref{fig:tbi-afmi}.
A similar signature can be reproduced already on the mean-field level, although
with only a kink instead of a cusp at the critical point. Results for the gap and the
mean-field order parameter are presented in the inset of
Fig.~\ref{fig:tbi-afm-gaps}(a). Figure~\ref{fig:tbi-afm-gaps}(a) also shows the
closing of the spin gap $\Ds$ at $\Uc$; the results
were obtained from the finite-size scaling shown in the inset of Fig.~\ref{fig:3DXY_gap}(a).

In Fig.~\ref{fig:tbi-afm-gaps}(b), we present numerical data for the
energy derivative
\begin{equation}\label{eq:dfdU}
    \frac{\partial F}{\partial U} 
    =    
    \las 
    \oh
    \sum_{\bm{i}} (c^{\dagger}_{\bm{i}} c^\nag_{\bm{i}} - 1 )^2
    \ras \,,
\end{equation}
corresponding to the expectation value of the interaction term or, equivalently,
the average double occupation, at $\lambda/t=0.1$. The continuous variation
of this quantity across $\Uc/t=4.95(5)$ suggests a continuous transition.

Having established the phase boundary of the magnetic transition at large $U/t$, 
we now consider the universality class. Given the remaining $U(1)$ spin
symmetry in the presence of spin-orbit coupling, the transition is expected to be in the 3D XY
universality class. An intuitive picture is based on local magnetic
moments, which already exist in the magnetically disordered phase for
$U>0$, and order at $\Uc$. The onset of phase coherence at $U=\Uc$
corresponds to a $U(1)$ symmetry breaking.  This scenario
is in accordance with the behavior of the spin gap $\Ds$ in
Fig.~\ref{fig:tbi-afm-gaps}. The excitons are massive in the disordered phase
($U<\Uc$), but condense in the ordered phase ($U\geq\Uc$) where
$\Ds=0$. 

The conjectured 3D XY universality can be tested using the zero-temperature,
finite-size scaling forms 
\begin{equation}\label{eq:scaling:S}
  \Sxy/N  = L^{-2\beta/\nu} f_1[(U-\Uc) L ^{1/\nu}]
\end{equation}
and
\begin{equation}\label{eq:scaling:D}
  \Ds/t  = L^{-z} f_2[(U-\Uc) L ^{1/\nu}]\,.
\end{equation}
Here $f_1$ and $f_2$ are dimensionless functions. The relevant
critical exponents for the 3D XY model are
$z=1$, $\nu=0.6717(1)$ and $\beta=0.3486(1)$.\cite{PhysRevB.74.144506}

Using the same value $\lambda/t=0.1$ as before, we show in
Fig.~\ref{fig:3DXY_SAFxy}(a) $L^{2\beta/\nu} \Sxy/N$ as a function of $U$ for different system
sizes $L$. If the scaling form Eq.~(\ref{eq:scaling:S}) with the critical
exponents of the 3D XY model is correct, we expect to see an intersect
of curves for different $L$ at $U=\Uc$. As shown in
Fig.~\ref{fig:3DXY_SAFxy}(a), this prediction is indeed borne out by the QMC data,
and we deduce $\Uc/t=4.96(4)$, in agreement with Fig.~\ref{fig:tbi-afmi}. Replotting
$L^{2\beta/\nu} \Sxy/N$ as a function of $L^{1/\nu}
(U-\Uc)/\Uc$ in Fig.~\ref{fig:3DXY_SAFxy}(b)
produces a clean scaling collapse onto a single
curve. Figure~\ref{fig:3DXY_SAFxy} hence demonstrates that the assumption of
3D XY behavior is fully consistent with the QMC data. 

Figure~\ref{fig:3DXY_gap} shows a similar analysis for the spin gap $\Ds$,
using the scaling form~(\ref{eq:scaling:D}). Although the statistical quality
of the data is not quite as good as for the structure factor, we again find
satisfactory scaling (in particular, there is no noticeable drift of the
intersect with increasing $L$) and the same $\Uc$ using the 3D XY critical exponents.

Based on the existence of a $U(1)$ spin symmetry throughout the TBI phase,
we expect the 3D XY behavior found at $\lambda/t=0.1$ to be generic for this transition,
in agreement with previous predictions.\cite{Hohenadler10,Gr.Xu.11,PhysRevLett.107.166806}
The finite-size corrections to the 3D XY scaling
behavior become more pronounced on approaching the possible multicritical
point, as verified explicitly for $\lambda/t=0.05$. According to Griset and
Xu,\cite{Gr.Xu.11} the observed 3D XY behavior at $\lambda>0$ suggests the 
absence of a chiral AF order-disorder transition inside the AFMI phase even for
$\lambda=0$.\cite{PhysRevB.84.024420}

\subsection{Spin liquid to antiferromagnet transition}\label{sec:qsl-afmi}

The refined phase boundary of the QSL phase shown in Fig.~\ref{fig:pd}
establishes the existence of a QSL--AFMI transition at finite $\lambda$,
in addition to the $\lambda=0$ transition studied before.\cite{Meng10}
Since the present work is concerned with the KMH model, we only consider
finite values $\lambda>0$ here. 

The simple picture of the magnetic transition as an ordering transition
of magnetic moments (or exciton condensation) discussed in the context
of the TBI--AFMI transition cannot straightforwardly be applied to
the QSL--AFMI transition. For example, a $Z_2$ spin liquid exhibits 
charge fractionalization, and therefore has no well-defined magnetic
modes. Fractionalization could lead to an unusually large anomalous
dimension.\cite{Is.Me.Ha.11} On the other hand, if the QSL phase was adiabatically
connected to a simple band insulator (without charge fractionalization), the
transition is again expected to be of the 3D XY type, similar to the TBI--AFMI transition. 

\begin{figure}[t]
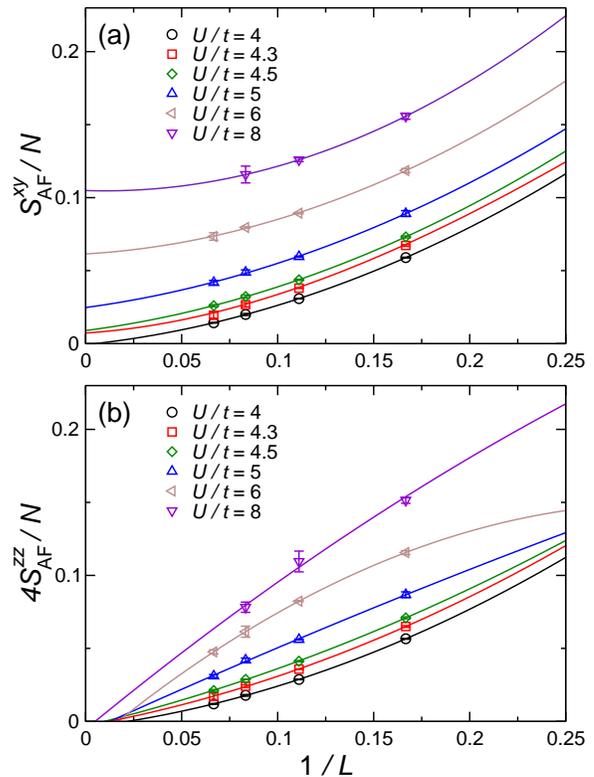

  \includegraphics[width=0.425\textwidth,clip]{fig_qsl-afmi_xy}\vspace*{0.5em}
  \includegraphics[width=0.425\textwidth,clip]{fig_qsl-afmi_zz}
  \caption{ \label{fig:xyorder} (Color online) 
    (a) Finite-size scaling of the rescaled magnetic structure factor
    $\Sxy/N$ defined in Eq.~(\ref{eq:SAF})  at
    $\lambda/t=0.0125$ for different values of $U/t$, across the QSL--AFMI
    transition.  The data  suggest a magnetic transition at
    $\Uc/t=4.3(2)$. Lines correspond to polynomial fits.
    (b) Same as in (a) but showing the longitudinal structure factor $\Szz$ defined in
    Eq.~(\ref{eq:SAFzz}). 
    In contrast to (a), there is no long-range order over the 
    range of $U/t$ values considered.
  }
\end{figure}

Due to the small size of the spin gap in the QSL phase, the
extrapolation of the order parameter~(\ref{eq:SAF}) to the thermodynamic
limit is much more delicate than for the TBI-AFMI transition. In particular,
a scaling analysis along the lines of Figs.~\ref{fig:3DXY_SAFxy} and~\ref{fig:3DXY_gap}
is not conclusive with the currently available system sizes.

We first address the question of longitudinal magnetic order.
The phase diagram presented by Yu \etal,\cite{Yu.Xie.Li.11} based on results
from the variational cluster approach, shows an extended region inside the AFMI phase
in which the authors claim that magnetic order exists both in the $xy$ plane and
in the $z$ direction. For $\lambda=0$, this region is argued to extend all the
way to the QSL--AFMI phase boundary, leading to a simultaneous onset of
transverse and longitudinal order at $\Uc$. At $\lambda>0$, Yu
\etal\cite{Yu.Xie.Li.11} find a transition from the TBI to an $xy$ ordered
AFMI at $\Uc$, and an onset of $z$ order at even larger values
of $U$. Hence, for $\lambda>0$, there would be an additional crossover (no symmetry breaking)
inside the AFMI phase. Whereas $z$ order is known to exist in the Hubbard
model ($\lambda=0$), this result is surprising in the light of the strong-coupling
picture mentioned above, in which antiferromagnetic correlations in the $z$
direction are frustrated by the interplay of hopping $t$ and spin-orbit coupling
$\lambda$. 

To clarify the situation, we use unbiased QMC simulations and calculate the 
transverse structure factor [Eq.~(\ref{eq:SAF})] as well as the longitudinal structure factor
%
%
%
\begin{align}\label{eq:SAFzz}
 \Szz
 &\equiv \sum_\alpha [\Szz]^{\alpha\alpha}\,,
 \\\nonumber
 [\Szz]^{\alpha\beta}
 &= \frac{1}{L^2}\sum_{\bm{r}\bm{r}'} (-1)^{\alpha} (-1)^{\beta} 
 \langle \Psi_0
 | S^{z}_{\bm{r}\alpha} S^{z}_{\bm{r}'\beta} | \Psi_0 \rangle\,,
\end{align}
at $\lambda/t=0.0125$. The results are shown in Fig.~\ref{fig:xyorder}. The
onset of transverse magnetic order is visible from the finite-size extrapolation of
$\Sxy/N$ depicted in Fig.~\ref{fig:xyorder}(a),
and the critical value $\Uc/t=4.3(2)$ is shown in the phase diagram in
Fig.~\ref{fig:pd}. However, as revealed by Fig.~\ref{fig:xyorder}(b),
there is no long-range order in the longitudinal direction even for large
values of $U/t=8$. We have carried out simulations down to $\lambda/t=0.002$,
where longitudinal order would be most favorable, but found no $z$ order for the
$U$ range shown in Fig.~\ref{fig:pd}. Hence, an extended region of
$z$ order as suggested by Ref.~\onlinecite{Yu.Xie.Li.11} does not exist, and
the phase diagram is instead given by Fig.~\ref{fig:pd}, with a very narrow,
possibly infinitesimal, region of coexisting longitudinal and transverse order near $\lambda=0$. The
discrepancy between our exact numerical results and those of the variational cluster approach is most
likely  a consequence of the very small cluster sizes used for the latter.
Although the strong-coupling picture with exchange constants $J,J'$
is not justified for intermediate $U$, the frustration in the $z$ direction
qualitatively explains the absence of longitudinal order found
numerically. The purely in-plane magnetic order agrees with
field-theory predictions for the KMH model.\cite{Gr.Xu.11}

\begin{figure}[t]
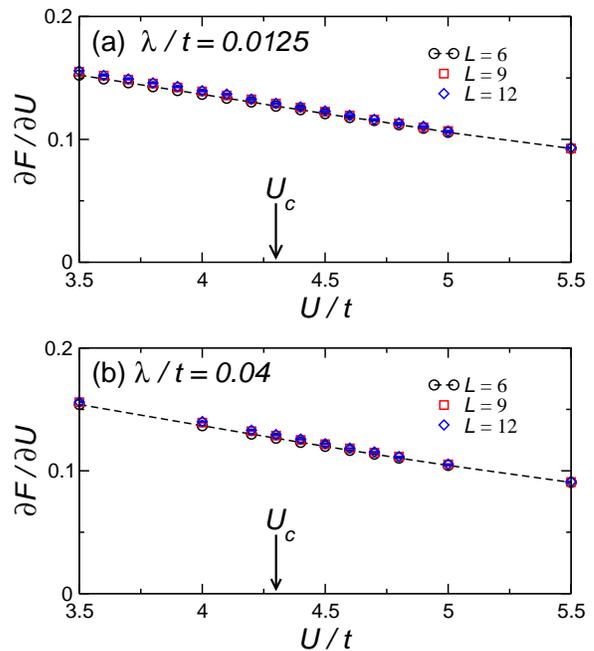

  \includegraphics[width=0.425\textwidth]{fig_qsl-afmi_dfdu_lambda0.0125.eps}\vspace*{0.75em}
  \includegraphics[width=0.425\textwidth]{fig_qsl-afmi_dfdu_lambda0.04.eps}
  \caption{\label{fig:qsl-tbi-docc} (Color online) Energy derivative
    with respect to $U$ [Eq.~(\ref{eq:dfdU})] across the QSL--AFMI transition at
    $\lambda/t=0.0125$ [$\Uc/t=4.3(2)$], and at $\lambda/t=0.04$
    [$\Uc/t=4.3(2)$], close to the multicritical point. There are no
    signs of a first-order transition.}
\end{figure}

Using field theory arguments, Griset and Xu\cite{Gr.Xu.11} suggested
the possibility that the QSL--AFMI transition could be first order.
To test this hypothesis, we show in Fig.~\ref{fig:qsl-tbi-docc}(a)
the energy derivative $ \partial F/\partial U$ [Eq.~(\ref{eq:dfdU})].
We do not find any sign of discontinuous behavior near $\Uc$, which suggests
that the transition is continuous. However, we cannot exclude
the possibility of a weakly first-order transition. We have also
calculated $ \partial F/\partial U$ at $\lambda/t=0.04$ (close to the
multicritical point) and found no signature of discontinuous behavior, see Fig.~\ref{fig:qsl-tbi-docc}(b).

\begin{figure}[ht]
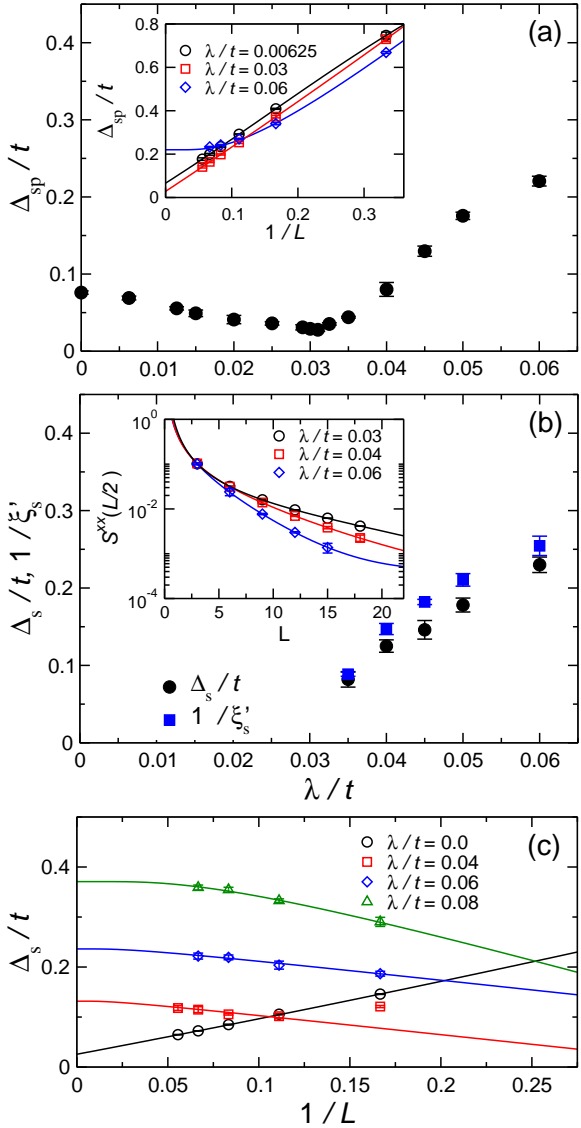

  \includegraphics[width=0.425\textwidth,clip]{fig_qsl-tbi_sp}\vspace*{0.5em}
  \includegraphics[width=0.425\textwidth,clip]{fig_qsl-tbi_spin}\vspace*{0.5em}
  \includegraphics[width=0.425\textwidth,clip]{fig_qsl-tbi_spin_scaling}
  \caption{\label{fig:qsl-tbi-sp} (Color online) (a) Single-particle gap 
    $\Dsp$  and (b) spin gap $\Ds$ and inverse spin correlation length
    $1/\xi_\text{s}'$ as a function of $\lambda$ at $U/t=4$, across the QSL--TBI
    transition. $\Dsp$ is obtained from finite-size scaling using
    Eq.~(\ref{eq:fit1}) for $\lambda/t\geq0.045$ and Eq.~(\ref{eq:fit2})
    for $\lambda/t<0.045$. The cusp defines the critical coupling $\lc/t=0.030(1)$.
    The inset in (a) shows the scaling for selected values of $\lambda/t$.    
    $\Ds$ is obtained from finite-size scaling using Eq.~(\ref{eq:fit1}), see
    (c). $\xi_\text{s}'$ is extracted from fits to the spin-spin correlation
    function defined in Eq.~(\ref{eq:spinspin}) at $r=L/2$,
    see inset in (b).}
\end{figure}

\subsection{Spin liquid to topological insulator transition}\label{sec:qsl-tbi}

A characteristic feature of both the QSL and the TBI phase is the absence of
broken symmetries. Therefore, the QSL--TBI transition cannot be tracked by
a local order parameter. In previous work,\cite{Hohenadler10} the
critical point $\lc$ at $U/t=4$ was determined from the behavior of the
single-particle gap. Here we discuss the underlying procedure in detail,
present new results with improved resolution of the critical point and based
on larger system sizes up to $L=18$, and refine the phase boundary by determining
the critical point at two other values of $U/t$. Moreover, we show numerical
results for the spin gap, and address the possibility of a first-order transition.

Figure~\ref{fig:qsl-tbi-sp}(a) shows the single-particle gap $\Dsp$ as
a function of $\lambda$ at $U/t=4$.  The data points are obtained from
extrapolation to the thermodynamic limit, as illustrated
in the inset. Deep in the TBI phase ($\lambda>\lc$), we use the fitting function
($\alpha=\text{sp}$, $\text{s}$)
\begin{equation}\label{eq:fit1}
  \Delta_\alpha(L)/t
  = 
  a + e^{-L/\xi_\alpha} (b/L   + c/L^2 )\,,
\end{equation}
with a correlation length $\xi_{\alpha}$. On approaching $\lc$ from
above, the correlation length set by the single-particle gap, $\xi_\text{sp}$,
increases and exceeds $L/2=9$ ($L=18$ being our largest system size) for
$\lambda/t\approx0.04$. For $\lambda/t$ smaller than 0.04, we use 
\begin{equation}\label{eq:fit2}
  \Delta_{\alpha}(L)/t =  a+ b/L   + c/L^2 \,.
\end{equation}
As a function of $\lambda$, the extrapolated single-particle gap in
Fig.~\ref{fig:qsl-tbi-sp}(a) initially
decreases when starting from the QSL at $\lambda=0$, reveals 
a cusp centered at $\lc/t=0.030(1)$, and increases
rather quickly with increasing $\lambda$ for $\lambda>\lc$. As in previous
work,\cite{Hohenadler10} we take the location of the cusp to define the
critical point $\lc$ of the QSL--TBI transition. We will argue below that
the data are consistent with a closing of the gap at $\lc$, and that the
cusp is a result of finite-size effects. For $\lambda/t\geq0.045$, the
larger system sizes now available result in larger values of $\Dsp$ compared
to previous work.\cite{Hohenadler10}

A similar analysis can be carried out for the spin gap $\Ds$ using
Eq.~(\ref{eq:fit1}) for $\lambda>\lc$. Similar to the SM--TBI transition
discussed above, the spin gap shows an unusual finite-size scaling inside the
TBI phase, as shown in Fig.~\ref{fig:qsl-tbi-sp}(c). The system sizes
required to see saturation (\ie, the magnetic correlation lengths) are
significantly larger at $U/t=4$ than at $U/t=2$,
cf. Fig.~\ref{fig:sm-tbi}(c). The extrapolated values of $\Ds$ for
$\lambda>\lc$ are shown in Fig.~\ref{fig:qsl-tbi-sp}(b). Comparing 
 $\Dsp$ and $\Ds$ [Figs.~\ref{fig:qsl-tbi-sp}(a) and \ref{fig:qsl-tbi-sp}(b)],
we see that in contrast to $U/t=2$ [Fig.~\ref{fig:sm-tbi}(a)] we have
$\Ds<2\Dsp$ in the TBI phase at $U/t=4$. The suppressed spin gap indicates
substantial particle-hole binding. In the QSL phase, the small values
of the spin gap make an accurate determination very challenging; $\Ds$ is
largest at $\lambda=0$, where it was previously determined as $\Ds/t=0.023(5)$.\cite{Meng10}

Figure~\ref{fig:qsl-tbi-sp}(b) reveals that the behavior of the spin gap for
$\lambda>\lc$ is very similar to that of $\Dsp$.  As a consistency check, we
also show the inverse spin correlation length $\xi_\text{s}$. Assuming the
form 
\begin{equation}\label{eq:spinspin}
  S^{xx}(r) = \langle S^x_{\bm{r}} S^x_{\bm{0}} \rangle   = e^{-r/\xi_\text{s}'} ( a/r   +  b/r^2)
\end{equation}
for the real-space transverse spin-spin correlation function, and taking the largest available
distance $r=L/2$ for each system size [see inset of
Fig.~\ref{fig:qsl-tbi-sp}(b)], the dependence of $1/\xi_\text{s}'$ on
$\lambda$ is in good agreement with $\Ds$ and $\Dsp$. We
only show the values $\xi_\text{s}'\leq L/2$.

For a noninteracting $Z_2$ TBI, there is a simple relation between the
excitation gaps in the single-particle sector and, \eg, in the spin and
particle-hole channels. In the presence of (strong) interactions, these
relations may be modified, and we indeed find $\Ds<2\Dsp$ in the TBI phase
above the QSL--TBI transition, as well as in the QSL phase at
$\lambda=0$.\cite{Meng10} The argument that $\Dsp$ has to close across a
transition that involves a change of the topological index\cite{HaKa10} holds
only for the noninteracting case. In general, it is not clear which
excitation gaps (one or more) close if the states on either side of the
transition are not adiabatically connected. For example, in the interacting
Haldane model,\cite{Va.Su.Ri.Ga.11} there is an exact degeneracy of the three
lowest states at the TBI to charge density wave transition. As a result, the
first and second excitation gaps ($E_1-E_0$ and $E_2-E_0$) close, but the
single-particle gap $\Dsp$ shows only a cusp at the critical point.

As argued in previous work,\cite{Hohenadler10} the results for $\Dsp$ are
consistent with a vanishing of the single-particle gap at $\lc$. Furthermore, the results in
Fig.~\ref{fig:qsl-tbi-sp} reveal that $\Dsp$ and $\Ds$ behave very similar on
approaching $\lc$ from above, and we may therefore expect to see a
simultaneous closing of $\Dsp$ and $\Ds$. Such a gap closing suggests
different Chern numbers for the TBI and QSL phases. Additionally,
the quick, almost linear opening of the gaps
for $\lambda>\lc$ is reminiscent of Fig.~\ref{fig:sm-tbi}(a) for the SM--TBI
transition, suggesting that a non-TBI phase (the QSL) exists at small values of
$\lambda$, and that a transition to the TBI phase takes place at $\lc>0$.
This picture confirms the expectation that the fully gapped QSL phase should
be stable under a small perturbation in the form of the spin-orbit term.

We attribute the small but nonzero values of the gaps at $\lc$
to finite-size effects. Although we used the same range of
system sizes (up to $L=18$) as for $\lambda=0$,\cite{Meng10} the larger
correlation lengths in the present case, especially in the spin channel, make
the analysis significantly harder.  On approaching $\lc$ from above, the
correlation lengths exceed the largest distance available on the clusters
used. If we consider only the data points for which the correlation lengths
fit on the largest system, \ie the range $\lambda/t\geq0.045$ in
Figs.~\ref{fig:qsl-tbi-sp}(a) and \ref{fig:qsl-tbi-sp}(b),
the functional form of $\Dsp$, $\Ds$ and $1/\xi_\text{s}'$ strongly suggests
a closing of the gaps very close to $\lc/t=0.030(1)$. The fact that the finite-size
scaled gaps in Figs.~\ref{fig:qsl-tbi-sp}(a) and \ref{fig:qsl-tbi-sp}(b) saturate at a finite
value close to $\lc$ is therefore likely to be a result of insufficiently large
system sizes. The latter require that we switch to the polynomial fitting
function~(\ref{eq:fit2}) close to $\lc$ for $\Dsp$, and do not permit
a reliable calculation of $\Ds$ or $\xi_\text{s}'$ close to $\lc$. The question if the gaps
close or not cannot be answered using approximate cluster
calculations,\cite{Yu.Xie.Li.11,Wu.Ra.Li.LH.11} because such methods are not
capable of describing a true QSL phase. 

To determine the shape of the QSL phase boundary, we have calculated the
single-particle gap for two other values of $U/t$. The critical
values $\lc$ are again defined by the location of the cusp in $\Dsp$.
We find $\lc/t=0.025(2)$ for $U/t=3.8$ and $\lc/t=0.032(2)$
for $U/t=4.1$.

\begin{figure}[t]
  \includegraphics[width=0.425\textwidth]{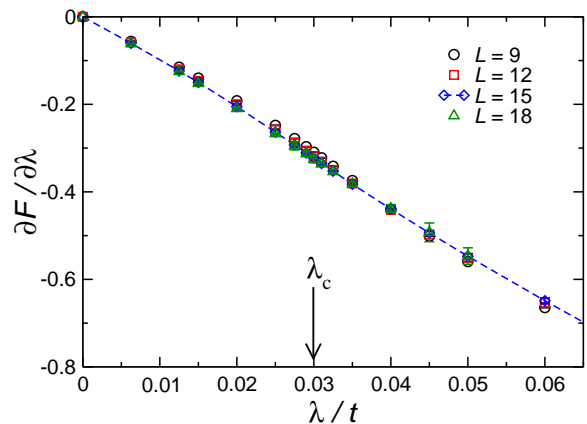}
  \caption{\label{fig:qsl-tbi-ener} (Color online) Energy derivative
    with respect to $\lambda$ [Eq.~(\ref{eq:dfdlambda})], across the QSL--TBI transition at
    $U/t=4$. There is no sign of a discontinuity at $\lc/t=0.030(1)$.}
\end{figure}

Recent theoretical work based on a $1/N$ expansion predicts the possibility
of a first-order QSL--TBI transition.\cite{Gr.Xu.11} To test this prediction,
we show in Fig.~\ref{fig:qsl-tbi-ener} the quantity
\begin{equation}\label{eq:dfdlambda}
  \frac{\partial F}{\partial \lambda} 
  =
  \las\, 
   i \! {\ensuremath \sum_{\llas\bm{i},\bm{j}\rras}} 
   \nu^\nag_{\bm{i}\bm{j}}
   c^{\dagger}_{\bm{i}}
   {\sigma}^z c^\nag_{\bm{j}}
  \ras
  \,,
\end{equation}
corresponding to the expectation value of the spin-orbit term in Eq.~(\ref{eq:H}).
For the range of system sizes, and on the very fine grid of $\lambda$ values,
there is no sign of a discontinuity. Again, we cannot rule out the possibility
of a weakly first-order transition.

\section{Conclusions and Outlook}\label{sec:conclusions}

Using exact quantum Monte Carlo simulations, we have obtained the phase
diagram of the Kane-Mele-Hubbard model (Fig.~\ref{fig:pd}). For
weak Hubbard interaction, the system is either a semimetal (SM) (at zero
spin-orbit coupling, $\lambda=0$) or a topological band insulator (TBI). The
latter is adiabatically connected to the noninteracting groundstate of the
Kane-Mele model, as evinced by the almost identical dependence of the single-particle and
spin gaps on $\lambda$. We have presented evidence for substantial
particle-hole binding in the TBI phase for small systems or large Hubbard
interaction. For intermediate Hubbard $U$, the
model supports a quantum spin liquid (QSL) phase at small $\lambda$ and a TBI phase at large
$\lambda$.  At large $U$, long-range magnetic order breaks time reversal
invariance, and the system becomes an antiferromagnetic Mott insulator (AFMI). 
In the presence of spin-orbit coupling, magnetic order is restricted to the
$xy$ plane. 

As previously suggested,\cite{Hohenadler10,Gr.Xu.11,PhysRevLett.107.166806}
the magnetic TBI--AFMI transition can be understood
as a condensation of magnetic excitons. A scaling analysis of the
magnetization and the spin gap provides clear evidence for the 3D XY
nature of the transition. The onset of long-range order coincides
with the closing of the spin gap, whereas the single-particle gap stays
finite but shows a cusp at the critical point.  In contrast to 
theoretical predictions, the corresponding transition between the QSL and the
AFMI appears to be continuous.

The QSL--TBI transition manifests itself as a cusp in the single-particle
and spin gap. The numerical data are compatible with a complete closing of
the gaps, but a definite conclusion is complicated by restrictions in
lattice sizes. The independently deduced inverse spin correlation length
is consistent with this picture, thereby suggesting
that the QSL and TBI phases are not adiabatically connected. Finally, we find
no sign of a predicted first-order transition.

There remain a number of interesting open issues, including a characterization of
the QSL phase, resolving the possible closing of the spin and single-particle
gap across the QSL--TBI transition, and the universality class of the QSL--AFMI
transitions both at $\lambda=0$ and $\lambda>0$. Understanding the
universality would provide important insight about the nature of the QSL
phase, including the possible existence of fractionalization. All 
these questions require significantly larger system sizes and
hence massively parallel computers and will be addressed in future work.

{\begin{acknowledgments}%
    We thank G.~Fiete, A.~Ruegg, C.~Varney and C.~Xu for useful
    discussions. We acknowledge support from the DFG Grants No.~FOR1162,
    SFB/TRR21 and WE 3639/2-1. This research was supported in part by the
    National Science Foundation under Grant No. NSF PHY0551164 and the NSF
    EPSCoR Cooperative Agreement No. EPS-1003897 with additional support from
    the Louisiana Board of Regents.
    Z.Y.M. acknowledges the hospitality of the Institute of Physics and KITPC at
    the Chinese Academy of Sciences. We are grateful to LRZ Munich, NIC
    J\"ulich, the J\"ulich Supercomputing Centre and HLR Stuttgart for
    generous allocation of computer time.
\end{acknowledgments}}


\end{document}